\documentclass[conference]{IEEEtran}
\IEEEoverridecommandlockouts

\usepackage{cite}
\usepackage{amsmath,amssymb,amsfonts,mathtools}
\usepackage{algorithmic}
\usepackage{graphicx}
\usepackage{textcomp}
\usepackage{booktabs}
\usepackage{multirow}
\usepackage{listings}
\usepackage{adjustbox}
\usepackage{url}
\usepackage{makecell}
\usepackage{bigdelim}
\usepackage{breqn}

\usepackage[T1]{fontenc}

\usepackage[labelformat=simple]{subcaption}

\usepackage[font=small,labelfont=bf]{caption}

\usepackage[colorlinks=true,linkcolor=blue,citecolor=blue]{hyperref}
\usepackage{cleveref}
\usepackage{tikz}

\renewenvironment{thebibliography}[1]{
  \begin{oldthebibliography}{#1}
    \setlength{\itemsep}{0em}
    \setlength{\parskip}{0em}
}
{
  \end{oldthebibliography}
}
\usepackage{setspace}

\allowdisplaybreaks

\def\BibTeX{{\rm B\kern-.05em{\sc i\kern-.025em b}\kern-.08em
    T\kern-.1667em\lower.7ex\hbox{E}\kern-.125emX}}

\newcommand{\phantomsubfigure}[1]{\begin{subfigure}[b]{0.0\textwidth}\phantomcaption\label{#1}\end{subfigure}}
\newcommand\circled[1]{\tikz[baseline=(char.base)]{\node[shape=circle,draw,inner sep=0.2pt] (char) {#1};}}

\begin{document}

\title{Efficient Streaming LLM for Speech Recognition}

\author{
     \IEEEauthorblockN{Junteng Jia, Gil Keren, Wei Zhou, Egor Lakomkin, Xiaohui Zhang, \\ Chunyang Wu, Frank Seide, Jay Mahadeokar, Ozlem Kalinli}
     \IEEEauthorblockA{Meta AI, USA
     \\ \texttt{juntengjia@meta.com}}
}

\maketitle

\begin{abstract}
Recent works have shown that prompting large language models with audio encodings can unlock speech recognition capabilities.
However, existing techniques do not scale efficiently, especially while handling long form streaming audio inputs --- not only do they extrapolate poorly beyond the audio length seen during training, but they are also computationally inefficient due to the quadratic cost of attention.  

In this work, we introduce SpeechLLM-XL, a linear scaling decoder-only model for streaming speech recognition.
We process audios in configurable chunks using limited attention window for reduced computation, and the text tokens for each audio chunk are generated auto-regressively until an EOS is predicted.
During training, the transcript is segmented into chunks, using a CTC forced alignment estimated from encoder output.
SpeechLLM-XL with 1.28 seconds chunk size achieves 2.7\%/6.7\% WER on LibriSpeech test clean/other, and it shows no quality degradation on long form utterances 10x longer than the training utterances.
\end{abstract}


\section{Introduction}
\begin{figure*}
    \centering
    \phantomsubfigure{fig:speech_llm_xl_a}
    \phantomsubfigure{fig:speech_llm_xl_b}
    \phantomsubfigure{fig:speech_llm_xl_c}
    \includegraphics[width=1.0\linewidth]{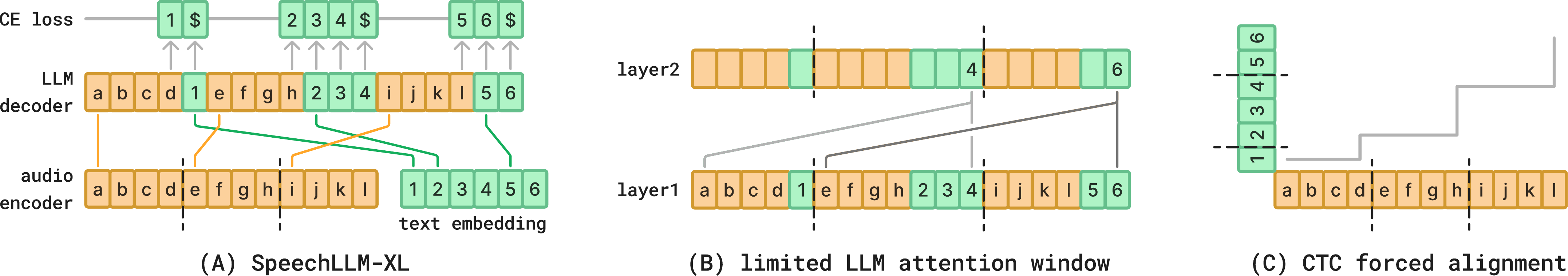}
    \caption{Overview of the proposed model. (A) SpeechLLM-XL consists of an audio encoder, a LLM decoder, and a text embedding layer. The audio sequence is processed in static-length chunks, and the resulting audio encodings (denoted as alphabet) are interleaved with text embedding (denoted as numbers) according to audio-text alignment, and the entire sequence is fed into the LLM. The model is trained for next-token-prediction to generate text tokens for each chunk, plus an EOS token \$ indicating the end-of-chunk. (B) We use a limited attention window in the LLM decoder to reduce computation. In this plot, the audio/text encodings in each chunk only attend to previous one chunk besides the current chunk (i.e. token 4 would attend to $\{a, b, c, d, 1, e, f, g, h, 2, 3, 4\}$). (C) During training, the audio-text alignment is computed using a CTC forced aligner to align audio encodings and text tokens.}
    \label{fig:speech_llm_xl}
    \vspace{-0.25in}
\end{figure*}
Recent studies have shown that a decoder-only large language model (LLM) pretrained on a large amount of text corpus can be adapted to understand multi-modal input (e.g. image and audio) by simply prompting the LLM with embeddings of the corresponding modality~\cite{Driess2023Palm,Zhu2023Minigpt,Team2023Gemini,Dubey2024Llama,Achiam2023Gpt}.
Specifically for automatic speech recognition (ASR), it has been established that a LLM can be finetuned to generate transcription when prompted with speech input~\cite{Fathullah2023PromptingLL,Wu2023On,Lakomkin2024End}.
In short, an audio encoder converts the input audio into a sequence of encodings, then the LLM auto-regressively predict all transcript tokens conditioned on the audio sequence.
%
%
%
This class of methods have achieved state-of-the-art (SoTA) accuracy on ASR benchmarks, and will be denoted as \textit{SpeechLLMs} in the following discussion.

Despite their success, a practical limitation of SpeechLLMs is that they are unsuitable for processing long form utterances.
First, SpeechLLMs have limited length extrapolation abilities, i.e., their performance significantly degrades when the audio length goes beyond the maximum audio length during training.
This is because SpeechLLMs are trained to predict an EOS to terminate decoding after all transcript tokens are generated, and they tend to terminate early when an utterance's transcript is longer than all the utterances the models have seen during training.
We will discuss this in detail in \cref{subsec:length_extrapolation}.
Second, even if we increase the length of training utterances to improve the accuracy on long form audios, a computational challenge is that per-token inference cost scales linearly with the audio length due to decoder LLM attention, and the overall inference cost scales quadratically with length.
Finally, deploying an ASR model to production system typically has a low latency requirement.
Since SpeechLLMs are non-streaming models that generate all transcript at once after an entire utterance has been received, the user perceived latency on long utterances would be much higher than short utterances.
This problem is further worsen by the quadratic-scaling inference cost.

In this work, we tackle these challenges on long form ASR with \textit{SpeechLLM-XL (extra long)}, which is a streaming model that scales linearly with audio length.
Our model consists of an audio encoder and a LLM decoder, and its key mechanism is audio chunking.
The input audio is segmented into static-length chunks, which implies each chunk could have a variable number of transcript tokens.
After, the input audio is processed chunk-by-chunk.
In particular, the encoding of the $k^{\texttt{th}}$ audio chunk is used to prompt the LLM decoder, which auto-regressively generate transcript tokens until an EOS is predicted; later, when the LLM is prompted with $(k+1)^{\texttt{th}}$ audio chunk, the previous audio chunks and decoded tokens are used as the LLM context.
We introduce two hyper-parameters to trade off accuracy verses latency and computation, 1) audio chunk size controls the model latency, 2) LLM context size controls the computational cost of the attention operations.
We will discuss those tradeoffs in detail in \cref{subsec:chunk_size,subsec:llm_context}.

SpeechLLM-XL is trained on paired audio/text data, and we use the audio/text alignment to segment the text transcript.
When the reference alignment is not available, we add an auxiliary CTC loss to the audio encoder and use a CTC forced aligner~\cite{Pratap2024Scaling} to align the encoder output and transcript.
We will analyze the quality of CTC forced alignment in \cref{subsec:ctc_forced_alignment}.

We conduct a throughout empirical study using LibriSpeech dataset, which shows SpeechLLM-XL successfully addresses all three aforementioned challenges on long form utterances.
First, to demonstrate its length extrapolation ability, we use a SpeechLLM-XL model trained on regular LibriSpeech training data to decode a concatenated version of \texttt{test-clean/other} utterances.
As discuss in \cref{subsec:length_extrapolation}, no quality degradation is observed on the concatenated long form utterances.
Second, to demonstrate its efficiency, we show in \cref{subsec:llm_context} that the attention window of the LLM decoder in SpeechLLM-XL can be aggressively reduced without hurting model accuracy.
Thus using a small static LLM context would reduce the overall inference cost from quadratic to linear in audio length, and enables efficient inference for long form utterances.
Finally, to demonstrate its streaming capability, we show in \cref{subsec:chunk_size} that a SpeechLLM-XL model with a small chunk size of 1.28 seconds gives comparable accuracy with the non-streaming SpeechLLM.
We also compare our model with other streaming baselines with similar latency in \cref{sec:additional_baselines}, and SpeechLLM-XL compares favorably against the existing methods.

\section{Background on SpeechLLM \label{sec:speech_llm}}
Given an input utterance $\mathbf{X} = (\mathbf{x}_1,\ldots,\mathbf{x}_T)$ where $\mathbf{x}_t \in \mathbb{R}^d$ denotes Log-Mel audio features, the goal is to predict the transcription $\mathbf{y} = (y_1,\ldots,y_U)$ where $y_u \in \mathcal{V}$ denotes a text token.
SpeechLLM estimates the conditional distribution of the transcription $\mathbf{y}$ given the audio $\mathbf{X}$:
\begin{align*}
P(\mathbf{y}|\mathbf{X}) \approx P_{\theta}(\mathbf{y}|\mathbf{X}) = \prod_{u=1}^{U} P_{\theta}(y_{u} | y_{1:u-1},\mathbf{X}) \cdot P_{\theta}(\$ | \mathbf{y},\mathbf{X})
\end{align*}
where \$ denotes the EOS token and $P_{\theta}$ is parametrized with a neural network.
For simplicity, we set $y_{U+1} = \$$ to extend the transcript $\mathbf{\tilde{y}} = (y_1,\ldots,y_U,\$)$ and rewrite:
\begin{align*}
P_{\theta}(\mathbf{y}|\mathbf{X}) = P_{\theta}(\mathbf{\tilde{y}}|\mathbf{X}) = \prod_{u=1}^{U+1} P_{\theta}(y_{u} | y_{1:u-1},\mathbf{X})
\end{align*}

During training, given a collection of paired audio/text data $\{(\mathbf{X}^{(n)}, \mathbf{y}^{(n)})\}_{n=1}^{N}$, the neural network is trained to maximize the conditional probability of the entire training set:
\begin{align*}
\theta &= \arg\max_{\theta} \prod_{n=1}^{N} P_{\theta}(\mathbf{\tilde{y}}^{(n)} | \mathbf{X}^{(n)}) \\
       &= \arg\min_{\theta} -\frac{1}{N} \sum_{n=1}^{N} \sum_{u=1}^{U+1} \log P_{\theta}(y_{u}^{(n)} | y_{1:u-1}^{(n)}, \mathbf{X}^{(n)})
\end{align*}
which is to minimize the averaged cross-entropy loss across all tokens including EOS.
During inference, given an input $\mathbf{X}$, the extended transcript $\mathbf{\tilde{y}}$ is generated auto-regressively using beam search, and EOS is removed for the final transcript $\mathbf{y}$.

\section{SpeechLLM-XL}
Given an audio/text pair $\mathbf{X},\mathbf{y}$, we segment $\mathbf{X}$ into chunks $(\mathbf{X}_{1} \ldots \mathbf{X}_{k} \ldots \mathbf{X}_{K})$ with chunk size $c$.
Assuming the audio/text alignment is known, we further segment $\mathbf{y}$ and define:
\begin{align*}
\mathbf{\tilde{y}} = (\mathbf{\tilde{y}}_{1}, \ldots, \mathbf{\tilde{y}}_{k}, \ldots \mathbf{\tilde{y}}_{K}) = (\mathbf{y}_{1}, \$, \ldots, \mathbf{y}_{k}, \$, \ldots, \mathbf{y}_{K}, \$)
\end{align*}
where $\mathbf{y}_{k}$ denotes the text in the $k^{\texttt{th}}$ chunk.
Then, we write the conditional probability of the entire sequence as:
\begin{align*}
P_{\theta}(\mathbf{\tilde{y}}|\mathbf{X}) = \prod_{k=1}^{K} P_{\theta}(\mathbf{\tilde{y}}_{k}|\mathbf{X}_{k}, \{\mathbf{\tilde{y}}_{1:k-1}, \mathbf{X}_{1:k-1}\})
\end{align*}
Note the probability of predicting $\mathbf{\tilde{y}}_{k}$ given the current chunk $\mathbf{X}_{k}$ depends on audio and transcripts of 
the previous chunks $\{\mathbf{\tilde{y}}_{1:k-1}, \mathbf{X}_{1:k-1}\}$.
To explicitly model this dependency, when computing LLM decoder self-attention on tokens representing the current chunk $k$, we attend to the cached attention values of the previous $b$ chunks; the total size of the LLM attention window is $b \times c$.
The entire architecture of our model is illustrated in \cref{fig:speech_llm_xl_a}.
In order to reduce the attention computation on long form utterances, we typically choose a small value $b$ to limit the LLM context.
For example, the model in \cref{fig:speech_llm_xl_b} use LLM context $b = 1$, which only covers the previous chunk, plus the current chunk.
%
%
We will study the impact of chunk size and LLM context size on accuracy in \cref{subsec:chunk_size,subsec:llm_context}.

So far, we have assumed the availability of ground truth alignment between audio and text.
When the alignment is not available, we use CTC forced alignment between audio encoding and text tokens to determine the token end time, as shown in \cref{fig:speech_llm_xl_c}.
Empirically, we found the CTC forced alignment is a good approximation and SpeechLLM-XL trained with CTC forced alignment gives similar accuracy as the reference alignment.
We will discuss the CTC forced alignment quality and its impact on overall model accuracy \cref{subsec:ctc_forced_alignment}.

During inference, our beam search decoding algorithm is similar to the alignment-synchronous search for 
neural transducers~\cite{Saon2020Alignment}, meaning in each decoding step, all hypotheses in the beam have the same number of tokens including the EOS.

\section{Experiments}
We benchmark our model performance and conduct ablation studies on the LibriSpeech dataset~\cite{Panayotov2015LibrispeechAA}, which consists of 960h training data sourced from English audio-books.
Audio features are 80 mel-spaced log filter-bank channels with a sliding window of 25ms at 10ms frame shift.
A time reduction layer stacks features with a stride of 4, and the
resulting 320-dimensional vectors form the input to the encoder.

Since the focus of this paper is on the SpeechLLM-XL architecture and its hyper-parameters, we keep the encoder, decoder and output vocabulary size constant across all experiments.
The audio encoder is a 20-layer streaming Emformer with 240ms of lookahead, 8 self-attention heads, embedding dimension 512, FFN dimension 2048, followed by a linear projection with dimension 768, totaling 107M parameters.
The decoder is a 12-layer LLM downsized from Llama2, with 12 self-attention heads, embedding dimension 768, FFN dimension 2048, totaling 92M parameters.
The output vocabulary is 4096 unigram SentencePieces estimated from the training transcripts, plus the CTC blank symbol.
The segment size of the Emformer encoder is the same as the audio segment size of SpeechLLM-XL.
The encoder is pretrained with CTC loss on LibriSpeech training data, and the decoder is pretrained on the transcript text corpus.
After, the encoder and decoder is combined and SpeechLLM-XL is trained on LibriSpeech training data with cross-entropy loss, plus a CTC auxiliary loss with weight 0.5.
We use a CTC forced aligner to estimate the audio/text alignment unless described otherwise.
We conduct a series of ablation studies on SpeechLLM-XL and denote the resulting models as $\circled{1} - \circled{\footnotesize{10}}$.
All models are trained with an Adam optimizer with a tri-stage learning rate 
schedule and a peak learning rate of $1 \times 10^{-4}$.
For each run, we select the best checkpoint based on the \texttt{dev-clean} WER, and we report the final WER on both \texttt{dev-clean/other} and \texttt{test-clean/other}.

\subsection{Chunk Size \label{subsec:chunk_size}}
We vary audio chunk size to explore the trade-off between latency and accuracy.
We fix the LLM context to 5.12s and consider a range of audio chunk size options from 2.56s to 0.32s, and we also include a non-streaming SpeechLLM\footnote{Following \cite{Fathullah2023PromptingLL}, the SpeechLLM baseline uses a non-streaming 
Conformer encoder consists of a convolutional frontend with stride 4 followed by 24 Conformer layer, totaling 110M parameters. It uses the same 92M parameters LLM decoder as SpeechLLM-XL.} as a baseline for comparison.
The results are summarized in \cref{tab:chunk_size}.

First, compared with the non-streaming SpeechLLM, the streaming model $\circled{1}$ with 2.56s chunk size gives better WER on \texttt{dev/test-clean}.
Looking into the WER breakdown on the \texttt{dev-clean} dataset, we find that SpeechLLM performs much worse on long utterances and the majorities of the mistakes are deletion errors.
In particular, on the 50 longest utterances in \texttt{dev-clean}, SpeechLLM gives a total WER of 16.4 and deletion rate of 13.9, while $\circled{1}$ gives a total WER of 2.2 and deletion rate of 0.1.
This confirms our hypothesis that SpeechLLM-XL is suitable for long utterances.
We will further compare the length extrapolation abilities of SpeechLLM-XL against SpeechLLM in \cref{subsec:length_extrapolation}.

Second, as we decrease the chunk size, WER increase on all evaluation datasets.
This is expected since reducing chunk size reduces the model latency.
We find a chunk size of 1.28s gives a good trade-off between latency and accuracy, and we will use this chunk size for the remaining of the paper.
\begin{table}[!t]
\caption{The effect of SpeechLLM-XL audio chunk size on model quality, where both the audio chunk size and LLM context size are measured in seconds. As the audio chunk size is reduced, the model latency is reduced at the cost of higher WER. The non-streaming SpeechLLM is included as a baseline.}
\label{tab:chunk_size}
\centering
\begin{tabular}{c|c|c||cc|cc}
\hline\hline
\multirow{2.4}{*}{Model} & \multirow{2.4}{*}{\shortstack{LLM\\context}} & \multirow{2.4}{*}{\shortstack{Chunk\\size}} & \multicolumn{2}{c|}{dev} & \multicolumn{2}{c}{test} \\
\cline{4-7}
 & & & clean & other & clean & other \\
\hline\hline
SpeechLLM                     & --                     & $\infty$  & 2.9    & 5.4    & 3.4    & 5.5    \\
\hline
$\circled{1}$                 & \multirow{4}{*}{5.12s} & 2.56s     & 2.4    & 6.8    & 2.5    & 6.5    \\
\cline{1-1} \cline{3-7}
$\circled{2}$                 &                        & 1.28s     & 2.5    & 6.9    & 2.7    & 6.7    \\
\cline{1-1} \cline{3-7}
$\circled{3}$                 &                        & 0.64s     & 2.7    & 7.3    & 2.9    & 7.4    \\
\cline{1-1} \cline{3-7}
$\circled{4}$                 &                        & 0.32s     & 3.0    & 7.9    & 3.1    & 7.8    \\
\hline\hline
\end{tabular}
\end{table}

\subsection{LLM Context \label{subsec:llm_context}}
Next, we study how the LLM context impact model quality.
We fix the audio chunk size to 1.28s and consider a range of LLM left context from $\infty$ (full context) to 0 (no context).
The results are summarized in \cref{tab:llm_context}.
First, we did not observe a considerable quality degradation as we reduce decoder context from $\infty$ to 1.28s.
This indicates SpeechLLM-XL works well with a small LLM context, thus the inference cost can be reduced from quadratic to linear in audio length without impacting accuracy.
However, when we fully remove all previous audio chunks from LLM context, we observe a notable quality degradation from $\circled{7}$ to $\circled{8}$.
This indicates a small LLM context is still necessary and naively segment the audio and decode each chunk separately does not work as well.

\begin{table}[!t]
\caption{The effect of LLM left context size on WER. There is little degradation in accuracy as we reduce the LLM context from $\infty$ to 1.28s, until we completely removed all the previous chunks from the LLM context.}
\label{tab:llm_context}
\centering
\begin{tabular}{c|c|c||cc|cc}
\hline\hline
\multirow{2.4}{*}{Model} & \multirow{2.4}{*}{\shortstack{LLM\\context}} & \multirow{2.4}{*}{\shortstack{Chunk\\size}} & \multicolumn{2}{c|}{dev} & \multicolumn{2}{c}{test} \\
\cline{4-7}
 & & & clean & other & clean & other \\
\hline\hline
$\circled{5}$ & $\infty$  & \multirow{6}{*}{1.28s} & 2.6    & 6.8    & 2.7    & 6.7    \\
\cline{1-2} \cline{4-7}
$\circled{2}$ &    5.12s  &                        & 2.5    & 6.9    & 2.7    & 6.7    \\
\cline{1-2} \cline{4-7}
$\circled{6}$ &    2.56s  &                        & 2.5    & 6.9    & 2.7    & 6.8    \\
\cline{1-2} \cline{4-7}
$\circled{7}$ &    1.28s  &                        & 2.5    & 6.8    & 2.7    & 6.8    \\
\cline{1-2} \cline{4-7}
$\circled{8}$ &       0   &                        & 2.7    & 7.4    & 3.0    & 7.2    \\
\hline\hline
\end{tabular}
\end{table}

\subsection{CTC Forced Alignment \label{subsec:ctc_forced_alignment}}
So far, we have used CTC forced alignment during training without examining alignment quality.
To this end, we compare the CTC forced alignment from the audio encoder verses the reference alignment from a chenone hybrid acoustic model~\cite{Le2019Senones}.
We measure the CTC forced alignment quality by the averaged \textit{alignment delay} and \textit{alignment delta} measured on LibriSpeech \texttt{dev-clean/other}.
The alignment delay is defined as the token end time difference averaged across all tokens $\frac{1}{U} \sum_{u=1}^{U} t(y_{u}) - \hat{t}(y_{u})$, where $t(y_{u})$ and $\hat{t}(y_{u})$ is the token end time for $y_{u}$ from CTC alignment and hybrid alignment, respectively.
The alignment delta is defined as the token end time absolute difference $\frac{1}{U} \sum_{u=1}^{U} |t(y_{u}) - \hat{t}(y_{u})|$.
To understand the impact of alignment quality on ASR accuracy, we also replicated $\circled{2}$ using the reference hybrid alignment for text segmentation.
The results are summarized in \cref{tab:ctc_forced_alignment}.

The token end time from CTC forced alignment is on-averaged 52ms ahead of the reference hybrid alignment, as indicated by the negative alignment delay.
Overall, the CTC alignment and the hybird alignment is very close, with an alignment delta of 63ms.
In terms of downstream ASR model quality, the SpeechLLM-XL model trained with CTC alignment slightly under-performs the model trained with hybrid alignment, likely due to the negative emission delay.
\begin{table}[!t]
\caption{The quality of CTC forced alignment compared against the reference hybrid alignment. The CTC alignment is on-averaged 52ms ahead of the hybrid alignment, as indicated by the negative alignment delay. The SpeechLLM-XL model trained with reference hybrid alignment slightly out-performs CTC forced alignment.}
\label{tab:ctc_forced_alignment}
\centering
\begin{tabular}{c|c|c|c||cc|cc}
\hline\hline
\multirow{2.4}{*}{Model} & \multirow{2.4}{*}{Align.} & \multirow{2.4}{*}{\shortstack{Align.\\delay}} & \multirow{2.4}{*}{\shortstack{Align.\\delta}} & \multicolumn{2}{c|}{dev} & \multicolumn{2}{c}{test} \\
\cline{5-8}
 & & & & clean & other & clean & other \\
\hline\hline
$\circled{2}$ &    CTC & -52ms & 63ms  & 2.5    & 6.9    & 2.7    & 6.7    \\
\cline{1-8}
$\circled{9}$ & hybrid & --    & --    & 2.4    & 6.7    & 2.7    & 6.5    \\
\hline\hline
\end{tabular}
\end{table}

\subsection{Length Extrapolation \label{subsec:length_extrapolation}}
One major advantage of SpeechLLM-XL over the baseline SpeechLLM is its length extrapolation ability.
To demonstrate this advantage, we conduct a length extrapolation experiment by repeating (and concatenating) each \texttt{test-clean/other} utterances multiple times.
We test the concatenated utterances on the SpeechLLM baseline and SpeechLLM-XL $\circled{2}$, and we summarize the results in \cref{tab:length_extrapolation}.
Unsurprisingly, SpeechLLM fails to extrapolate beyond the audio length the original model was trained on.
The WER more than tripled on concatenated \texttt{test-clean/other} utterances 2x longer than the utterances seen during training.
In contrast, SpeechLLM-XL extrapolates perfectly beyond the training length, even yielding a slightly lower WER on 2x/10x concatenated \texttt{test-other} utterances.

The length extrapolation ability is important for production deployment because 1) it allows the ASR model to be more robust to edge case input, 2) it allows training on segmented short utterances for better batching and training efficiency, while still maintaining model quality on long form utterances.

\begin{table}[!t]
\caption{The length extrapolation ability of SpeechLLM and SpeechLLM-XL, both of which are trained on regular LibriSpeech utterances.
SpeechLLM significantly degrade when tested on concatenated utterances that are 2x of the training length.
SpeechLLM-XL extrapolates to 10x of the training length with no quality degradation.}
\label{tab:length_extrapolation}
\centering
\begin{tabular}{c||cc|cc|cc}
\hline\hline
\multirow{2.4}{*}{Model} & \multicolumn{2}{c|}{test (1x)} & \multicolumn{2}{c|}{test (2x)} & \multicolumn{2}{c}{test (10x)} \\
\cline{2-7}
 & clean & other & clean & other & clean & other \\
\hline\hline
SpeechLLM     & 3.4 & 5.5 & 15.8 & 15.1 & 80.6 & 80.0 \\
\hline
$\circled{2}$ & 2.7 & 6.7 &  2.7 & 6.6 & 2.7 & 6.5 \\
\hline\hline
\end{tabular}
\end{table}

\section{Additional Baselines and Related Works \label{sec:additional_baselines}}
We compare SpeechLLM-XL with some additional baselines\footnote{Pushing for SoTA WER on LibriSpeech often requires self-supervised pre-training, pseudo labeling, or LM shallow fusion, and is not the purpose of this work, therefore we do not consider baselines with those modeling tricks.}.
The results are summarized\footnote{The LAS model is estimated to be over 270M parameters~\cite{Seide2024Speech}. Chunked AED did not specify model size. ReaLLM considers on two model size 82M and 7B, we present the numbers on their 82M model, which has a lower WER. CTC-prompt-LM is perhaps the smallest, with a total of 52M parameters.} in \cref{tab:additional_baselines}.

First, we consider some classic baselines.
LAS with Spec-Augment~\cite{Park2019Specaugment} is widely used non-streaming baseline without additional training tricks, where the encoder is a BiLSTM and decoder is a RNN.
SpeechLLM-XL out-performs this model despite being streaming and smaller-in-size.
For the streaming baselines, we trained a CTC model and a Transducer model.
The CTC model is an Emformer with the same look ahead and chunk size as the SpeechLLM-XL encoder, but with 28 layers and hidden dimension 768, totaling 246M parameters; the Transducer model further adds a predictor-joiner network on top, totalling 256M parameters.
SpeechLLM-XL performs notably better than both the CTC and the Transducer models.

\begin{table}[!t]
\caption{SpeechLLM-XL is competitive when compared to other streaming ASR models with similar latency and training recipe.}
\label{tab:additional_baselines}
\centering
\begin{tabular}{c|c|c||cc|cc}
\hline\hline
\multirow{2.4}{*}{Model} & \multirow{2.4}{*}{\shortstack{Look\\ahead}} & \multirow{2.4}{*}{\shortstack{Chunk\\size}} & \multicolumn{2}{c|}{dev} & \multicolumn{2}{c}{test} \\
\cline{4-7}
 & & & clean & other & clean & other \\
\hline\hline
LAS              & --    & --    & --     & --    & 2.8 & 6.8    \\
\hline
CTC              & 0.24s & 1.28s & 3.4    & 8.7   & 3.6 & 8.8    \\
\hline
Transducer       & 0.24s & 1.28s & 2.9    & 7.6   & 3.0 & 7.7    \\
\hline \hline
SpeechLLM-XL     & 0.24s & 1.28s & \textbf{2.5}    & 6.9    & \textbf{2.7}     & \textbf{6.7}    \\
\hline \hline
Chunked AED      & 0.30s & 1.20s & --     & \textbf{6.7}    & --     & \textbf{6.7}            \\
\hline
ReaLLM    & 0.96s & 1.92s & 2.7    & 7.6    & 3.0    & 7.4    \\
\hline
CTC-prompt-LM    & 0.64s & 0.64s & --     & --     & 3.2    & 7.9    \\
\hline\hline
\end{tabular}
\end{table}

Second, many cross-attention based encoder-decoder (AED) models were proposed for streaming ASR~\cite{Chiu2017Monotonic,Hsiao2020Online,Tsunoo2021Streaming,Zeyer2023Monotonic,Barrault2023Seamlessm4t,Radford2023Robust,Zeineldeen2024Chunked}.
Chunked AED~\cite{Zeineldeen2024Chunked} is the most recent and also most relevant to our work, and it explores a similar chunking idea for long form ASR.
SpeechLLM-XL gives a similar WER when compared against Chunked AED under a similar decoder chunk size and encoder look ahead.
The main difference is that SpeechLLM-XL uses self-attention based decoder-only model architecture, which makes it easier to integrate with a pretrained LLM.

Finally, although there has been a lot of works on using decoder-only LLMs for speech-processing~\cite{Fathullah2023PromptingLL,Wu2023On,Lakomkin2024End,Fathullah2023AudioChatLlamaTG,Chang2023Speechprompt,Zhang2023Speechgpt,Rubenstein2023Audiopalm,Arora2023Integrating,Maiti2024Voxtlm} and efficient streaming LLMs~\cite{Xiao2023Efficient,Miao2023Towards,Munkhdalai2024Leave,Huang2023Advancing,Zhou2024Survey,Dai2019Transformer,Ma2024Megalodon}, leveraging decoder-only models for streaming ASR is an under-explored area.
One recent work is ReaLLM~\cite{Seide2024Speech}, where each encoder output frame is used to prompt the LLM to generate its transcript (possibly empty), followed by a blank symbol.
ReaLLM scales quadratically with audio length, and it adds one additional blank prediction per frame during inference.
To reduce the attention computation and the number of blank predictions, the authors use a large encoder stride of 24 in their experiments.
SpeechLLM-XL performs notably better than ReaLLM, despite using a smaller encoder look ahead and decoder chunk size.
Another relevant work is \cite{Tsunoo2024Decoder}, which uses blank-filtered CTC encoder frames from each chunk to prompt the LM (therefore denoted as \textit{CTC-prompt-LM}).
In contrast to our work, they find CTC forced alignment works poorly on their model, and they instead use random prefix prompt for model training.
Most importantly, rather than having the decoder to output an EOS to move to next chunk, they use a joint CTC and decoder score fusion beam search to determine the end of chunk.
As such, the max decoding length for each chunk is upper-bounded by the CTC predictions.
SpeechLLM-XL out-performs CTC-prompt-LM by a considerable margin.

\bibliographystyle{IEEEbib}
\bibliography{refs}

\begin{thebibliography}{10}

\bibitem{Driess2023Palm}
Danny Driess, Fei Xia, Mehdi~SM Sajjadi, Corey Lynch, Aakanksha Chowdhery, Brian Ichter, Ayzaan Wahid, Jonathan Tompson, Quan Vuong, Tianhe Yu, et~al.,
\newblock ``Palm-e: An embodied multimodal language model,''
\newblock {\em arXiv preprint arXiv:2303.03378}, 2023.

\bibitem{Zhu2023Minigpt}
Deyao Zhu, Jun Chen, Xiaoqian Shen, Xiang Li, and Mohamed Elhoseiny,
\newblock ``Minigpt-4: Enhancing vision-language understanding with advanced large language models,''
\newblock {\em arXiv preprint arXiv:2304.10592}, 2023.

\bibitem{Team2023Gemini}
Gemini Team, Rohan Anil, Sebastian Borgeaud, Yonghui Wu, Jean-Baptiste Alayrac, Jiahui Yu, Radu Soricut, Johan Schalkwyk, Andrew~M Dai, Anja Hauth, et~al.,
\newblock ``Gemini: a family of highly capable multimodal models,''
\newblock {\em arXiv preprint arXiv:2312.11805}, 2023.

\bibitem{Dubey2024Llama}
Abhimanyu Dubey, Abhinav Jauhri, Abhinav Pandey, Abhishek Kadian, Ahmad Al-Dahle, Aiesha Letman, Akhil Mathur, Alan Schelten, Amy Yang, Angela Fan, et~al.,
\newblock ``The llama 3 herd of models,''
\newblock {\em arXiv preprint arXiv:2407.21783}, 2024.

\bibitem{Achiam2023Gpt}
Josh Achiam, Steven Adler, Sandhini Agarwal, Lama Ahmad, Ilge Akkaya, Florencia~Leoni Aleman, Diogo Almeida, Janko Altenschmidt, Sam Altman, Shyamal Anadkat, et~al.,
\newblock ``Gpt-4 technical report,''
\newblock {\em arXiv preprint arXiv:2303.08774}, 2023.

\bibitem{Fathullah2023PromptingLL}
Yassir Fathullah, Chunyang Wu, Egor Lakomkin, J.~Jia, Yuan Shangguan, Ke~Li, Jinxi Guo, Wenhan Xiong, Jay Mahadeokar, Ozlem Kalinli, Christian Fuegen, and Michael~L. Seltzer,
\newblock ``Prompting large language models with speech recognition abilities,''
\newblock {\em ICASSP 2024 - 2024 IEEE International Conference on Acoustics, Speech and Signal Processing (ICASSP)}, pp. 13351--13355, 2023.

\bibitem{Wu2023On}
Jian Wu, Yashesh Gaur, Zhuo Chen, Long Zhou, Yimeng Zhu, Tianrui Wang, Jinyu Li, Shujie Liu, Bo~Ren, Linquan Liu, and Yu~Wu,
\newblock ``On decoder-only architecture for speech-to-text and large language model integration,''
\newblock in {\em Workshop of Automatic Speech Recognition and Understanding}. IEEE, December 2023.

\bibitem{Lakomkin2024End}
Egor Lakomkin, Chunyang Wu, Yassir Fathullah, Ozlem Kalinli, Michael~L Seltzer, and Christian Fuegen,
\newblock ``End-to-end speech recognition contextualization with large language models,''
\newblock in {\em ICASSP 2024-2024 IEEE International Conference on Acoustics, Speech and Signal Processing (ICASSP)}. IEEE, 2024, pp. 12406--12410.

\bibitem{Pratap2024Scaling}
Vineel Pratap, Andros Tjandra, Bowen Shi, Paden Tomasello, Arun Babu, Sayani Kundu, Ali Elkahky, Zhaoheng Ni, Apoorv Vyas, Maryam Fazel-Zarandi, et~al.,
\newblock ``Scaling speech technology to 1,000+ languages,''
\newblock {\em Journal of Machine Learning Research}, vol. 25, no. 97, pp. 1--52, 2024.

\bibitem{Saon2020Alignment}
George Saon, Zolt{\'a}n T{\"u}ske, and Kartik Audhkhasi,
\newblock ``Alignment-length synchronous decoding for rnn transducer,''
\newblock {\em ICASSP 2020 - 2020 IEEE International Conference on Acoustics, Speech and Signal Processing (ICASSP)}, pp. 7804--7808, 2020.

\bibitem{Panayotov2015LibrispeechAA}
Vassil Panayotov, Guoguo Chen, Daniel Povey, and Sanjeev Khudanpur,
\newblock ``Librispeech: An asr corpus based on public domain audio books,''
\newblock {\em 2015 IEEE International Conference on Acoustics, Speech and Signal Processing (ICASSP)}, pp. 5206--5210, 2015.

\bibitem{Le2019Senones}
Duc Le, Xiaohui Zhang, Weiyi Zheng, Christian F{\"u}gen, Geoffrey Zweig, and Michael~L Seltzer,
\newblock ``From senones to chenones: Tied context-dependent graphemes for hybrid speech recognition,''
\newblock in {\em 2019 IEEE Automatic Speech Recognition and Understanding Workshop (ASRU)}. IEEE, 2019, pp. 457--464.

\bibitem{Seide2024Speech}
Frank Seide, Morrie Doulaty, Yangyang Shi, Yashesh Gaur, Junteng Jia, and Chunyang Wu,
\newblock ``Speech reallm--real-time streaming speech recognition with multimodal llms by teaching the flow of time,''
\newblock {\em arXiv preprint arXiv:2406.09569}, 2024.

\bibitem{Park2019Specaugment}
Daniel~S Park, William Chan, Yu~Zhang, Chung-Cheng Chiu, Barret Zoph, Ekin~D Cubuk, and Quoc~V Le,
\newblock ``Specaugment: A simple data augmentation method for automatic speech recognition,''
\newblock {\em arXiv preprint arXiv:1904.08779}, 2019.

\bibitem{Chiu2017Monotonic}
Chung-Cheng Chiu and Colin Raffel,
\newblock ``Monotonic chunkwise attention,''
\newblock {\em arXiv preprint arXiv:1712.05382}, 2017.

\bibitem{Hsiao2020Online}
Roger Hsiao, Dogan Can, Tim Ng, Ruchir Travadi, and Arnab Ghoshal,
\newblock ``Online automatic speech recognition with listen, attend and spell model,''
\newblock {\em IEEE Signal Processing Letters}, vol. 27, pp. 1889--1893, 2020.

\bibitem{Tsunoo2021Streaming}
Emiru Tsunoo, Yosuke Kashiwagi, and Shinji Watanabe,
\newblock ``Streaming transformer asr with blockwise synchronous beam search,''
\newblock in {\em 2021 IEEE Spoken Language Technology Workshop (SLT)}. IEEE, 2021, pp. 22--29.

\bibitem{Zeyer2023Monotonic}
Albert Zeyer, Robin Schmitt, Wei Zhou, Ralf Schl{\"u}ter, and Hermann Ney,
\newblock ``Monotonic segmental attention for automatic speech recognition,''
\newblock in {\em 2022 IEEE Spoken Language Technology Workshop (SLT)}. IEEE, 2023, pp. 229--236.

\bibitem{Barrault2023Seamlessm4t}
Lo{\"\i}c Barrault, Yu-An Chung, Mariano~Cora Meglioli, David Dale, Ning Dong, Paul-Ambroise Duquenne, Hady Elsahar, Hongyu Gong, Kevin Heffernan, John Hoffman, et~al.,
\newblock ``Seamlessm4t-massively multilingual \& multimodal machine translation,''
\newblock {\em arXiv preprint arXiv:2308.11596}, 2023.

\bibitem{Radford2023Robust}
Alec Radford, Jong~Wook Kim, Tao Xu, Greg Brockman, Christine McLeavey, and Ilya Sutskever,
\newblock ``Robust speech recognition via large-scale weak supervision,''
\newblock in {\em International conference on machine learning}. PMLR, 2023, pp. 28492--28518.

\bibitem{Zeineldeen2024Chunked}
Mohammad Zeineldeen, Albert Zeyer, Ralf Schl{\"u}ter, and Hermann Ney,
\newblock ``Chunked attention-based encoder-decoder model for streaming speech recognition,''
\newblock in {\em ICASSP 2024-2024 IEEE International Conference on Acoustics, Speech and Signal Processing (ICASSP)}. IEEE, 2024, pp. 11331--11335.

\bibitem{Fathullah2023AudioChatLlamaTG}
Yassir Fathullah, Chunyang Wu, Egor Lakomkin, Ke~Li, Junteng Jia, Shangguan Yuan, Jay Mahadeokar, Ozlem Kalinli, Christian Fuegen, and Michael Seltzer,
\newblock ``Audiochatllama: Towards general-purpose speech abilities for llms,''
\newblock {\em Proceedings of the 2024 Conference of the North American Chapter of the Association for Computational Linguistics: Human Language Technologies (Volume 1: Long Papers)}, 2023.

\bibitem{Chang2023Speechprompt}
Kai-Wei Chang, Yu-Kai Wang, Hua Shen, Iu-thing Kang, Wei-Cheng Tseng, Shang-Wen Li, and Hung-yi Lee,
\newblock ``Speechprompt v2: Prompt tuning for speech classification tasks,''
\newblock {\em arXiv preprint arXiv:2303.00733}, 2023.

\bibitem{Zhang2023Speechgpt}
Dong Zhang, Shimin Li, Xin Zhang, Jun Zhan, Pengyu Wang, Yaqian Zhou, and Xipeng Qiu,
\newblock ``Speechgpt: Empowering large language models with intrinsic cross-modal conversational abilities,''
\newblock {\em arXiv preprint arXiv:2305.11000}, 2023.

\bibitem{Rubenstein2023Audiopalm}
Paul~K Rubenstein, Chulayuth Asawaroengchai, Duc~Dung Nguyen, Ankur Bapna, Zal{\'a}n Borsos, F{\'e}lix de~Chaumont Quitry, Peter Chen, Dalia~El Badawy, Wei Han, Eugene Kharitonov, et~al.,
\newblock ``Audiopalm: A large language model that can speak and listen,''
\newblock {\em arXiv preprint arXiv:2306.12925}, 2023.

\bibitem{Arora2023Integrating}
Siddhant Arora, Hayato Futami, Yosuke Kashiwagi, Emiru Tsunoo, Brian Yan, and Shinji Watanabe,
\newblock ``Integrating pretrained asr and lm to perform sequence generation for spoken language understanding,''
\newblock {\em arXiv preprint arXiv:2307.11005}, 2023.

\bibitem{Maiti2024Voxtlm}
Soumi Maiti, Yifan Peng, Shukjae Choi, Jee-weon Jung, Xuankai Chang, and Shinji Watanabe,
\newblock ``Voxtlm: Unified decoder-only models for consolidating speech recognition, synthesis and speech, text continuation tasks,''
\newblock in {\em ICASSP 2024-2024 IEEE International Conference on Acoustics, Speech and Signal Processing (ICASSP)}. IEEE, 2024, pp. 13326--13330.

\bibitem{Xiao2023Efficient}
Guangxuan Xiao, Yuandong Tian, Beidi Chen, Song Han, and Mike Lewis,
\newblock ``Efficient streaming language models with attention sinks,''
\newblock {\em arXiv preprint arXiv:2309.17453}, 2023.

\bibitem{Miao2023Towards}
Xupeng Miao, Gabriele Oliaro, Zhihao Zhang, Xinhao Cheng, Hongyi Jin, Tianqi Chen, and Zhihao Jia,
\newblock ``Towards efficient generative large language model serving: A survey from algorithms to systems,''
\newblock {\em arXiv preprint arXiv:2312.15234}, 2023.

\bibitem{Munkhdalai2024Leave}
Tsendsuren Munkhdalai, Manaal Faruqui, and Siddharth Gopal,
\newblock ``Leave no context behind: Efficient infinite context transformers with infini-attention,''
\newblock {\em arXiv preprint arXiv:2404.07143}, 2024.

\bibitem{Huang2023Advancing}
Yunpeng Huang, Jingwei Xu, Zixu Jiang, Junyu Lai, Zenan Li, Yuan Yao, Taolue Chen, Lijuan Yang, Zhou Xin, and Xiaoxing Ma,
\newblock ``Advancing transformer architecture in long-context large language models: A comprehensive survey,''
\newblock {\em arXiv preprint arXiv:2311.12351}, 2023.

\bibitem{Zhou2024Survey}
Zixuan Zhou, Xuefei Ning, Ke~Hong, Tianyu Fu, Jiaming Xu, Shiyao Li, Yuming Lou, Luning Wang, Zhihang Yuan, Xiuhong Li, et~al.,
\newblock ``A survey on efficient inference for large language models,''
\newblock {\em arXiv preprint arXiv:2404.14294}, 2024.

\bibitem{Dai2019Transformer}
Zihang Dai, Zhilin Yang, Yiming Yang, Jaime Carbonell, Quoc~V Le, and Ruslan Salakhutdinov,
\newblock ``Transformer-xl: Attentive language models beyond a fixed-length context,''
\newblock {\em arXiv preprint arXiv:1901.02860}, 2019.

\bibitem{Ma2024Megalodon}
Xuezhe Ma, Xiaomeng Yang, Wenhan Xiong, Beidi Chen, Lili Yu, Hao Zhang, Jonathan May, Luke Zettlemoyer, Omer Levy, and Chunting Zhou,
\newblock ``Megalodon: Efficient llm pretraining and inference with unlimited context length,''
\newblock {\em arXiv preprint arXiv:2404.08801}, 2024.

\bibitem{Tsunoo2024Decoder}
Emiru Tsunoo, Hayato Futami, Yosuke Kashiwagi, Siddhant Arora, and Shinji Watanabe,
\newblock ``Decoder-only architecture for streaming end-to-end speech recognition,''
\newblock {\em arXiv preprint arXiv:2406.16107}, 2024.

\end{thebibliography}

\end{document}